\documentclass[longauth,letter]{aa} 

\usepackage[]{graphicx} 
\usepackage{longtable}
\usepackage{lscape}
\usepackage{txfonts}
\usepackage{array}

\usepackage{rotating}
\usepackage{color}
\usepackage{url}
\usepackage{natbib}
\usepackage{float}
\usepackage{textcomp,gensymb}
\usepackage[utf8]{inputenc}
\usepackage{amsmath} 

\makeatother

\newcolumntype{M}{>{\scriptsize\centering\arraybackslash}m{1.6cm}}
\newcolumntype{N}{>{\scriptsize\centering\arraybackslash}m{0.6cm}}
\begin{document}

   \title{The evolution of luminous red nova \object{AT~2017jfs} in \object{NGC~4470}\thanks{Table 1 is only available in electronic form
at the CDS via anonymous ftp to cdsarc.u-strasbg.fr (130.79.128.5)
or via http://cdsweb.u-strasbg.fr/cgi-bin/qcat?J/A+A/}}
 \titlerunning{LRN \object{AT~2017jfs} in \object{NGC~4470}}
 
   \author{A. Pastorello\inst{1}\thanks{ \email{andrea.pastorello@inaf.it}}
          \and
           T.-W. Chen\inst{2}
          \and
          Y.-Z. Cai\inst{1,3}
          \and
          A. Morales-Garoffolo\inst{4}
          \and
          Z. Cano\inst{5}
          \and
          E. Mason\inst{6}
          \and
          E.~A. Barsukova\inst{7}
          \and
          S.~Benetti\inst{1}
          \and
          M.~Berton\inst{8,9}
          \and
          S.~Bose\inst{10}
           \and
          F.~Bufano\inst{11}
          \and
          E.~Callis\inst{12}
          \and
          G.~Cannizzaro\inst{13,14}
          \and
          R.~Cartier\inst{15}
          \and
          Ping~Chen\inst{10}
          \and
          Subo~Dong\inst{10}
          \and
          S.~Dyrbye\inst{16}
          \and
          N.~Elias-Rosa\inst{17,18}
           \and
          A. Fl\"ors\inst{19,20,21} 
          \and
          M.~Fraser\inst{12}
          \and
          S. Geier\inst{22,23}
          \and
          V.~P. Goranskij\inst{24}
          \and
          D.~A.~Kann\inst{5} 
          \and
          H.~Kuncarayakti\inst{8,25}
          \and
          F.~Onori\inst{26}
          \and
          A. Reguitti\inst{27}
          \and
          T. Reynolds\inst{25}
          \and
          I.~R.~Losada\inst{28,29}
          \and
          A.~Sagu\'es~Carracedo\inst{30}
          \and
          T.~Schweyer\inst{2}
          \and
          S.~J.~Smartt\inst{31}
          \and
          A.~M.~Tatarnikov\inst{24}
          \and
          A.~F. Valeev\inst{7,32}
          \and
          C. Vogl\inst{20,21}
           \and
          T.~Wevers\inst{33}
           \and
          A.~de~Ugarte~Postigo\inst{5,34}
          \and
          L.~Izzo\inst{5}
          \and
          C.~Inserra\inst{35}
          \and
          E.~Kankare\inst{25}
          \and
          K.~Maguire\inst{31,36}
          \and
          K.~W. Smith\inst{31}
          \and
          B.~Stalder\inst{37}
          \and
          L.~Tartaglia\inst{38}
          \and
          C.~C.~Th\"one\inst{5}
          \and
          G.~Valerin\inst{3}
          \and
          D.~R.~Young\inst{31}
}

\institute{
    INAF - Osservatorio Astronomico di Padova, Vicolo dell'Osservatorio 5, I-35122 Padova, Italy 
         \and
    Max-Planck-Institut f\"ur Extraterrestrische Physik, Giessenbachstra{\ss}e 1, 85748, Garching bei M\"unchen,, Germany
         \and
    Dipartimento di Fisica e Astronomia, Universit\`a di Padova, Vicolo dell'Osservatorio 3, I-35122 Padova, Italy
        \and
    Department of Applied Physics, University of C\'adiz, Campus of Puerto Real, E-11510 C\'adiz, Spain 
         \and
    Instituto de Astrof\'isica de Andaluc\'ia (IAA-CSIC), Glorieta de la Astronom\'ia s/n, E-18008, Granada, Spain         
         \and
    INAF - Osservatorio Astronomico di Trieste, Via G.B. Tiepolo 11, I-34143 Trieste, Italy
         \and
    Special Astrophysical Observatory, Russian Academy of Sciences, Nizhnij Arkhyz, Karachai-Cherkesia, 369167 Russia
         \and
   Finnish Centre for Astronomy with ESO (FINCA), University of Turku, Quantum, Vesilinnantie 5, FI-20014 University of Turku, Finland
      \and
  Aalto University Mets{\"a}hovi Radio Observatory, Mets{\"a}hovintie 114, FI-02540 Kylm{\"a}l{\"a}, Finland 
         \and
    Kavli Institute for Astronomy and Astrophysics, Peking University, Yi He Yuan Road 5, Hai Dian District, Beijing 100871, China
       \and
  INAF - Osservatorio Astrofisico di Catania, Via Santa Sofia 78, I-95123 Catania, Italy
        \and
    School of Physics, O'Brien Centre for Science North, University College Dublin, Belfield Dublin 4, Ireland
         \and
    SRON, Netherlands Institute for Space Research, Sorbonnelaan 2, 3584CA, Utrecht, The Netherlands
         \and
    Department of Astrophysics/IMAPP, Radboud University, P.O. Box 9010, 6500 GL Nijmegen, The Netherlands
         \and
    Cerro Tololo Inter-American Observatory, National Optical Astronomy Observatory, Casilla 603, La Serena, Chile
         \and
    Nordic Optical Telescope, Apartado 474, E-38700 Santa Cruz de La Palma, Santa Cruz de Tenerife, Spain
         \and
    Institute of Space Sciences (ICE, CSIC), Campus UAB, Cam\'{i} de Can Magrans s/n, 08193 Cerdanyola del Vall\`es (Barcelona), Spain
         \and
    Institut d’Estudis Espacials de Catalunya (IEEC), c/Gran Capit\`a 2-4, Edif. Nexus 201, 08034 Barcelona, Spain
         \and
    European Southern Observatory, Karl-Schwarzschild-Stra{\ss}e 2, 85748 Garching bei M\"unchen, Germany 
         \and
    Max-Planck-Institut f\"ur Astrophysik, Karl-Schwarzschild-Stra{\ss}e 1, 85748 Garching bei M\"unchen, Germany 
         \and
    Physik-Department, Technische Universit\"at M\"unchen, James-Franck-Stra{\ss}e 1, 85748 Garching bei M\"unchen, Germany
         \and
    Gran Telescopio Canarias (GRANTECAN), Cuesta de San Jos\'e s/n, E-38712, Bre\~na Baja, La Palma, Spain 
         \and
    Instituto de Astrof\'isica de Canarias, V\'ia L\'actea s/n, E-38200, La Laguna, Tenerife, Spain 
         \and
    Sternberg Astronomical Institute, Lomonosov Moscow University, Universitetsky Ave. 13, 119992 Moscow, Russia
         \and
    Tuorla Observatory, Department of Physics and Astronomy, University of Turku, FI-20014 Turku, Finland
           \and
    Istituto di Astrofisica e Planetologia Spaziali (INAF), via del Fosso del Cavaliere 100, I-00133 Roma, Italy
        \and
    Departamento de Ciencias F\'{i}sicas, Universidad Andr\'{e}s Bello, Santiago, Chile
         \and
    Nordita, KTH Royal Institute of Technology and Stockholm University,  SE-10691 Stockholm, Sweden
         \and
    Department of Astronomy, AlbaNova University Center, Stockholm University, SE-10691 Stockholm, Sweden
         \and
    The Oskar Klein Centre, Department of Physics, Stockholm University, AlbaNova, SE-10691 Stockholm, Sweden
         \and
    Astrophysics Research Centre, School of Mathematics and Physics, Queen's University Belfast, Belfast BT7 1NN, UK
         \and
    Kazan Federal University, Kremlevskaya 18, 420008 Kazan, Russia
         \and
    Institute of Astronomy, University of Cambridge, Madingley Road, Cambridge CB3 0HA, United Kingdom
         \and
    Dark Cosmology Centre, Niels Bohr Institute, University of Copenhagen, Juliane Maries Vej 30, 2100 Copenhagen, Denmark
        \and  
    School of Physics $\&$ Astronomy, Cardiff University, Queens Buildings, The Parade, Cardiff CF24 3AA, UK
         \and
    School of Physics, Trinity College Dublin, Dublin 2, Ireland
         \and
    LSST, 950 North Cherry Avenue, Tucson, AZ 85719, USA
         \and
    The Oskar Klein Centre, Department of Astronomy, Stockholm University, AlbaNova, SE-10691, Stockholm, Sweden
}

   \date{Received Month dd, 20yy; accepted Month dd, 20yy}

  \abstract
   {We present the results of our photometric and spectroscopic follow-up of the intermediate-luminosity optical transient \object{AT~2017jfs}. At peak, the object reaches an absolute magnitude of $M_g=-15.46\pm0.15$ mag and a bolometric luminosity of $5.5 \times 10^{41}$ erg s$^{-1}$. Its light curve has the double-peak shape typical of luminous red novae (LRNe), with a narrow first peak bright in the blue bands, while the second peak is longer-lasting and more luminous in the red and near-infrared (NIR) bands. During the first peak, the spectrum shows a blue continuum with narrow emission lines of H and Fe II. During the second peak, the spectrum becomes cooler, resembling that of a K-type star, and the emission lines are replaced by a forest of narrow lines in absorption. About 5 months later, while the optical light curves are characterized by a fast linear decline, the NIR ones show a moderate rebrightening, observed until the transient disappears in solar conjunction. At these late epochs, the spectrum becomes reminiscent of that of M-type stars, with prominent molecular absorption bands. The late-time properties suggest the formation of some dust in the expanding common envelope or an IR echo from foreground pre-existing dust. We propose that the object is a common-envelope transient, possibly the outcome of a merging event in a massive binary, similar to \object{NGC4490-2011OT1}.}


   \maketitle

\section{Introduction}

Red Novae (RNe) form a family of optical transients spanning an enormous range of luminosities. This
includes faint objects with absolute peak magnitudes $M_V$ from $-4$ to $-6.5$ mag, 
such as \object{OGLE 2002-BLG-360} \citep{tyl13} and \object{V1309~Sco} \citep{mas10,tyl11}, intermediate-luminosity
events ($M_V\gtrsim-10$ mag) like V838~Mon \citep{mun02,gor02,kim02,cra03}, and relatively luminous
objects such as \object{NGC4490-2011OT1} \citep{smi16}, that can reach $M_V\approx-15$ mag.
Objects brighter than $M_V=-10$ mag are conventionally named luminous red novae\footnote{The alternative 
naming `luminous red variable' was also used in the past \citep[e.g.,][]{mar99}.} \citep[LRNe; for a review, see][and references therein]{pasto19}.

Although the physical processes triggering these outbursts have been debated, there is growing evidence
that RNe and their more luminous counterparts are produced by the coalescence of stars with different
masses following a common-envelope phase \citep[e.g.,][]{koc14,pej16,pej17,mac17,mac18}. 
In particular, the inspiralling motion of the secondary was revealed by the long-term monitoring of \object{V1309~Sco} \citep{tyl11}.

The recent discovery of LRNe suggests that common envelope ejections and/or merging events may also happen
in more massive close binary systems \citep{smi16,mau18}, with major implications for the evolution of
the resulting merger. In this context, here we report the results of our follow-up
campaign of a LRN recently discovered in the galaxy \object{NGC~4470}: \object{AT~2017jfs}.

\section{\object{AT 2017jfs},  its host galaxy, and reddening} \label{host}

\object{AT~2017jfs}\footnote{The object is known by multiple survey designations, including \object{Gaia17dkh}, \object{PS17fqp},
\object{ATLAS18aat}.} was discovered by Gaia on 2017 December 26.13 UT \citep[MJD =  58113.13,][]{del17} at a Gaia $G$-band magnitude
17.17 $\pm$ 0.20. No source was detected at the transient position on 2017 November 30 down to a limiting
magnitude of 21.5. The transient was located at  $\alpha = 12^{h}29^{m}37\fs79$ and
 $\delta = +07\degr49\arcmin35\farcs18$ (equinox J2000.0), in the almost face-on, early-type spiral galaxy \object{NGC~4470}. 

The source was tentatively classified by the extended-Public ESO Spectroscopic Survey for 
Transient Objects \citep[ePESSTO;][]{sma15} as a type-IIn supernova (SN IIn) or a SN impostor
\citep{buf18}, and for this reason it was designated with a SN name (SN~2017jfs). In this paper, we show 
it to be a LRN, hence we adopt the label \object{AT~2017jfs}.

   \begin{figure}
   \includegraphics[width=0.37\textwidth,angle=270]{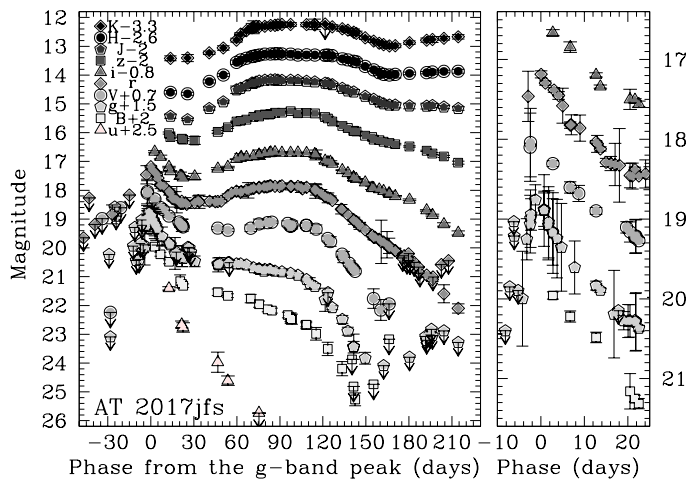}
      \caption{Left: Optical and NIR light curves of \object{AT~2017jfs}. The phases are calculated from the first $g$-band peak, on MJD = 58114.8 $\pm$ 1.8. Only the most significant detection limits are shown. 
Right: $B$, $V$, $g$, $r$ and $i$ light curves during the first (blue) peak.}
         \label{fig1}
   \end{figure}

The distance to \object{NGC~4470} is somewhat controversial, and the NASA/IPAC Extragalactic Database (NED)\footnote{\it https://ned.ipac.caltech.edu/} 
gives a number of discrepant estimates based on the Tully-Fisher 
method, ranging from 11.6 to 34.6 Mpc, with an average value of 18.76 $\pm$ 6.6 Mpc (corresponding to a distance 
modulus $\mu = 31.25 \pm 0.70$ mag). Given the uncertainty in the Tully-Fisher estimates, we prefer to adopt a kinematic distance  $d = 35.2 \pm 2.7$ Mpc
(corrected for Virgo Infall and estimated adopting a standard cosmology with $H_0$ = 73 km s$^{-1}$ Mpc$^{-1}$), hence $\mu$ = 32.73 $\pm$ 0.15 mag.
This estimate also agrees with the distance $d \sim$ 34.7 Mpc reported by \citet{kol17}.

The Galactic line-of-sight reddening is modest, $E(B-V)$ = 0.022 mag \citep{sch11}. Our early spectra do not have high signal to noise ratios (S/Ns),
and therefore the host galaxy reddening cannot be well constrained. However, prominent absorption features of Na~ID
are not visible, suggesting a modest host galaxy reddening contribution.
Later spectra with higher S/N  show only a narrow Na~ID in absorption centered at 5884\AA~(rest wavelength), hence
likely a feature intrinsic to \object{AT~2017jfs}. For this reason, we adopt $E(B-V)$ = 0.022 mag as the total reddening towards \object{AT~2017jfs}.

\section{Photometric evolution} \label{photometry}

The follow-up campaign started soon after the classification of \object{AT~2017jfs}, and continued for about 7 months. 
Photometry data were reduced following standard prescriptions \citep[see, e.g.,][]{cai18}, using the SNOoPy package \citep{cap14}.
The magnitudes are listed in Table 1, available at the CDS, which contains the following information: Column 1 lists the date of the observation, Column 2 lists the MJD, Columns 3 to 11 give the optical and near-infrared (NIR) magnitudes, and Column 12 reports a numeric code for the instrumental configuration. The multi-band light curves
are shown in Fig. \ref{fig1}. 
The Sloan-$u$ light curve shows a monotonic decline after maximum, with an average rate of 6.5 $\pm$ 0.9 mag (100 d)$^{-1}$.
The photometric evolution in the other bands is somewhat different.
The $g$-band maximum is constrained to MJD = 58114.8 $\pm$ 1.8 (2017 December 27.8 UT) through a low-order polynomial fit
(at $g = 17.35 \pm 0.02$ mag, hence $M_g = -15.46 \pm 0.15$ mag).

The $g$-band light curve  has a rise time to maximum of about 4\,d, followed by a 
rapid decline (6.5 $\pm$ 0.2 mag (100 d)$^{-1}$) until $\sim$50\,d. The light curve
follows a plateau-like evolution until $\sim$110\,d when it begins a faster decline
(7.0 $\pm$ 0.4 mag (100 d)$^{-1}$) that lasts until it has faded below the detection threshold.
The evolution in the Johnson $B$ and $V$ bands is similar to the $g$ band, although the $V$-band light curve shows a 
low-contrast second peak, broader than the early one.

The light curve is remarkably different in the red and NIR bands. The transient reaches a peak at $r = 17.19 \pm 0.05$ mag 
on MJD = 58115.6, followed by a fast decline (6.3 $\pm$ 0.3 mag (100 d)$^{-1}$) lasting three weeks and reaching a minimum at
$r = 18.32 \pm 0.07 $ mag. Subsequently, from about 50~d after maximum, the $r$-band luminosity rises again and reaches a second maximum 
on MJD = 58209.0, at $r = 17.68 \pm 0.03$ mag. This second peak is much broader than the early one. Later on,
from 110 to 200 d after the first peak, the $r$-band light curve declines with a rate of 3.90 $\pm$ 0.04 mag (100 d)$^{-1}$. 
The $i$-band light curve is very similar, with the two peaks reaching comparable luminosities.

The NIR light curves have a second maximum, brighter than the early one. As an additional feature, GROND \citep{gre08} observations reveal a moderate rebrightening of the NIR light 
curves from $\sim$170 to 220 d. Although we do not have very late spectroscopic observations to support this (Sect. \ref{spectroscopy}), a late-time NIR luminosity excess can be associated with the
formation of new dust or IR echoes, occasionally observed in LRNe at late phases \citep[see, e.g.,][]{ban15,ext16}. The late NIR brightening may also be a consequence of the transition to
the  brown (L-type) supergiant stage, as happened for V838~Mon \citep{eva03,mun07}, although this scenario does not comfortably explain the late blue-shift of the H$\alpha$ emission observed in 
the late spectra of \object{AT~2017jfs} (see Sect. \ref{spectroscopy}).

\section{Spectral evolution} \label{spectroscopy}

\begin{figure}
   \includegraphics[width=0.49\textwidth]{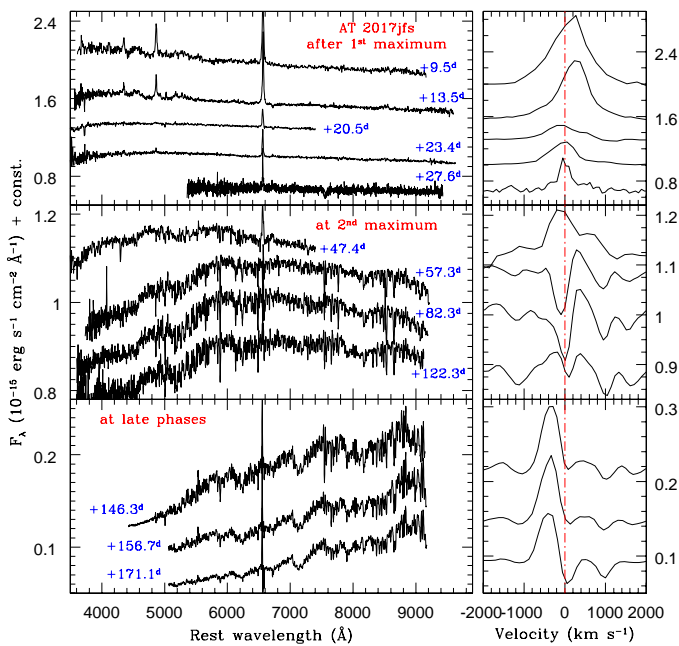}
   \caption{Left: Spectral evolution of \object{AT~2017jfs} soon after the first blue maximum (top), around the second, red maximum (center), and at late phases (bottom). Right: Expanded image of the H$\alpha$ region in the velocity space. The vertical line marks the rest wavelength position of H$\alpha$ at the same phases as the left panels (after correcting for the host galaxy redshift $z=0.007809$ as reported in NED). Only the highest-S\textbackslash N\ spectra are shown. The absorptions visible at velocities 0 and +1000 km s$^{-1}$ in late spectra can be partly due to an over-subtraction of the unresolved $H\alpha$ and [N II] $\lambda$6583 from the host galaxy background.}
              \label{fig2}%
    \end{figure}

\begin{figure}
   \centering
   \includegraphics[width=8.7cm]{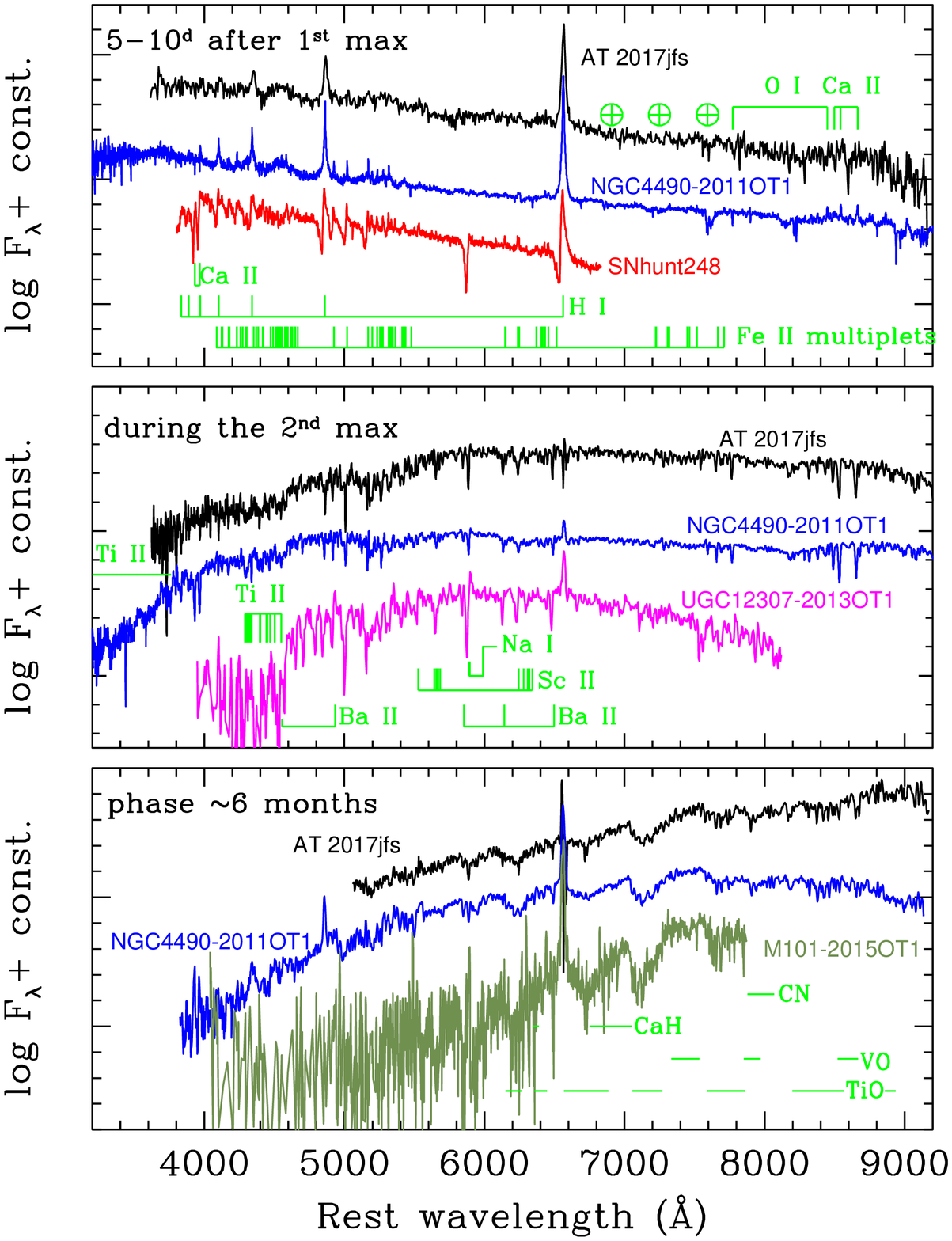}
   \caption{Line identification in the spectra of \object{AT~2017jfs} and a few comparison LRNe at three representative phases: a few days after the blue peak (top panel), around the red peak (mid panel),
and at late phases (about 5-6 months after the blue peak; bottom panel). The spectra of the comparison objects are taken from \protect\citet{pasto19}, \protect\citet{bla17}, and \protect\citet{kan15}. The identification of the molecular bands is performed following \protect\citet{kir91}, \protect\citet{val98}, \protect\citet{mar99}, and \protect\citet{bar14}.}
              \label{fig3}%
    \end{figure}

We collected 14 epochs of optical spectroscopy, spanning about six months of the evolution of \object{AT~2017jfs}. Information on the instrumental configurations is given in Table \ref{tab}.
Our spectral  sequence of \object{AT~2017jfs} is presented in Fig. \ref{fig2}, while the comparison with a few LRNe at similar epochs and the line identification are shown in  Fig. \ref{fig3}. We remark that the transient lies in a crowded region of \object{NGC~4470}, rich in nearby sources. As a consequence, the late spectra
show some contamination from host galaxy lines.

The spectral evolution of AT~2017jfs follows a three-phase behavior, as observed in other extra-galactic LRNe.
In particular, we note a remarkable similarity with \object{NGC4490-2011OT1} \citep{smi16,pasto19} at all phases.
At early epochs (until 3-4 weeks after the first $g$-band peak) the spectrum of \object{AT~2017jfs} shows a blue continuum, dominated by prominent H lines in emission, with a Lorentzian profile
and a full width at half maximum velocity $v_{FWHM} \sim 700$ km s$^{-1}$ (corrected for spectral resolution).
In this period, the temperature inferred from a black-body fit to the spectral continuum, $T_{BB}$, decreases from about $7800\pm700$ K (in the +9.5d spectrum) to $6000\pm600$ K (in the +23.4 d spectrum).
Emission lines from a number of Fe II multiplets are also detected, along with O~I. The Ca~II NIR triplet is also identified in emission, while the H$\&$K feature, which is usually prominent in absorption
in other LRNe (see Fig. \ref{fig3}, top panel), is marginally detected in \object{AT~2017jfs}. 

With time, the continuum becomes redder and the spectrum experiences an evident metamorphosis. During the second peak, from  $\sim$50 d to 4 months, the red spectrum
($T_{bb}=4300\pm700$ K at 82.3 d) is dominated by a forest of metal lines in absorption. The H lines become much weaker, showing now a P Cygni profile (see Fig. \ref{fig2}, left panel), although an over-subtraction of the narrow H$\alpha$ emission from a nearby H~II region may affect the apparent strength of the absorption.
The spectrum in this phase is reminiscent of intermediate-type stars (e.g., late G to K types). We identify a number of metal lines (from Fe~II, Ti~II, Sc~II and Ba~II multiplets), 
along with the Na I 5889,~5895 \AA~doublet (see Fig. \ref{fig3}, mid panel). The O~I and the NIR Ca~II triplet are now seen in absorption. The velocity of the narrow Fe~II lines deduced from the
wavelengths of absorptions is about 450 km s$^{-1}$. Some of the absorption lines visible at this stage are likely due to neutral metals, in particular 
Fe~I at red wavelengths.

From about 5 months after maximum (hence during the steep, late luminosity decline; see Sect. \ref{photometry}), the spectrum changes again, becoming much
redder ($T_{bb} \approx 2950 \pm 150$ K at 157 d) and closer to that of an M-type star. H$\alpha$ is now mostly in emission, with an evident blue-shift of 
its peak (by about 400 km s$^{-1}$; see Fig. \ref{fig2}, right), and a deep absorption at the rest velocity. While an over-subtraction of the contaminant H~II 
region can be responsible in part for this absorption, the strong blue-shift of the emission is real. The H$\alpha$ profile is similar to that of LRN \object{NGC4490-2011OT1} 
about 200 days after maximum \citep{smi16}, whose bluest emission peak was shifted by $-$280 km s$^{-1}$. 
Following \citet{smi16}, the development of a blue-shifted component in emission would be consistent with an expanding, shock-heated line-forming region, possibly
with aspherical symmetry. A peculiar geometry or, alternatively, the formation of dust hiding the rear emitting region (or both) may explain the late H$\alpha$ profile 
for both \object{NGC4490-2011OT1} and \object{AT~2017jfs}. Higher-resolution spectra and a wider temporal monitoring would help in discriminating the different scenarios.

As observed in similar transients \citep[e.g.,][]{pasto19}, the late-time spectrum is also characterized by broad absorption bands.
The features are generally identified as being due to molecules, in particular TiO and VO, although CN and CaH are not ruled out (see Fig. \ref{fig3}, top panel).

\section{Evolution of the temperature and the radius}

The double-peaked light-curve evolution and the major spectroscopic transition from an SN~IIn-like spectrum to that of a late-type
star are two remarkable properties of LRNe \citep{pasto19}. The photometric information in particular can be used to study how the
spectral energy distribution (SED), the effective temperature, and the photospheric radius evolve with time.

    \begin{figure}
{\includegraphics[width=0.24\textwidth]{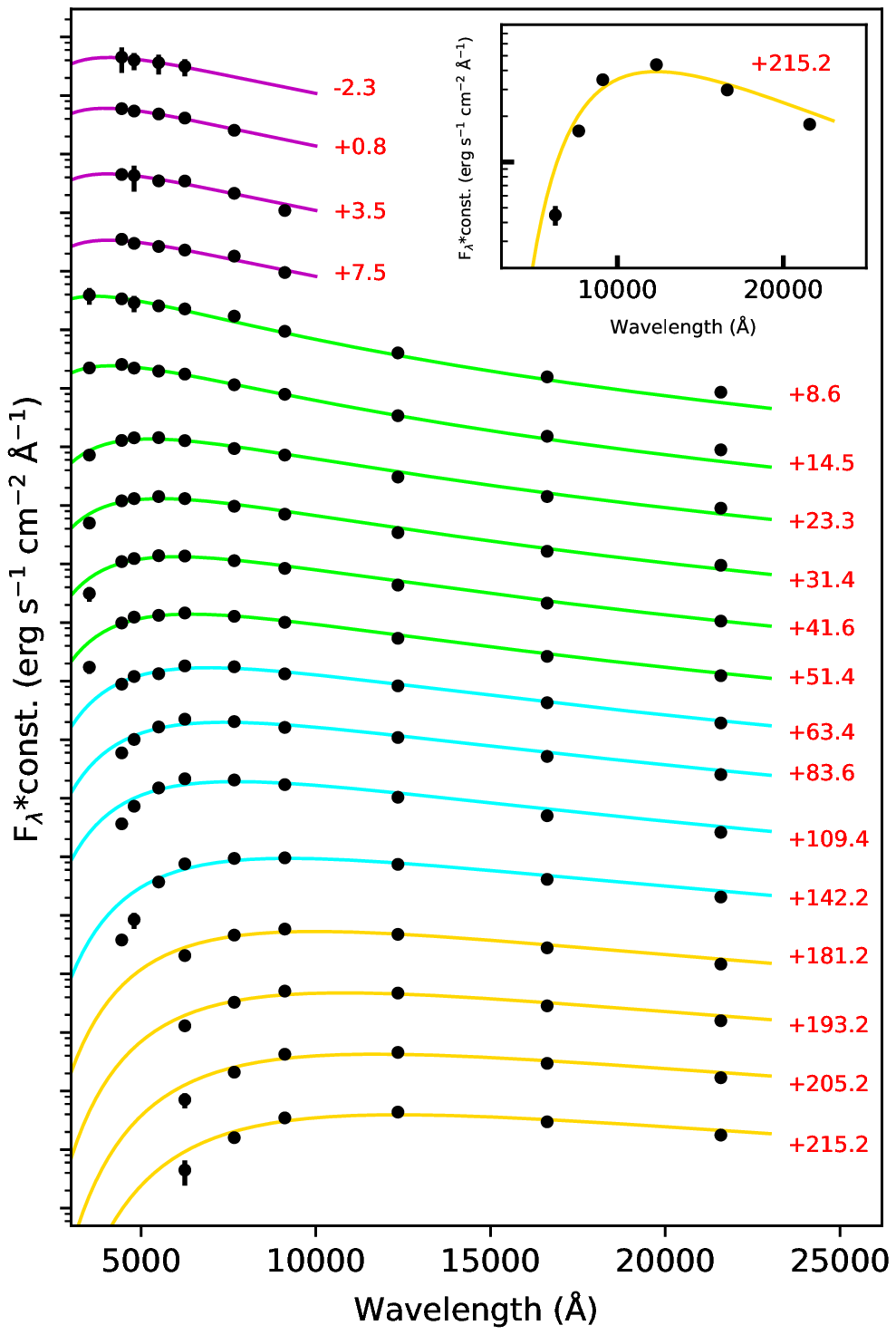}
 \includegraphics[width=0.24\textwidth]{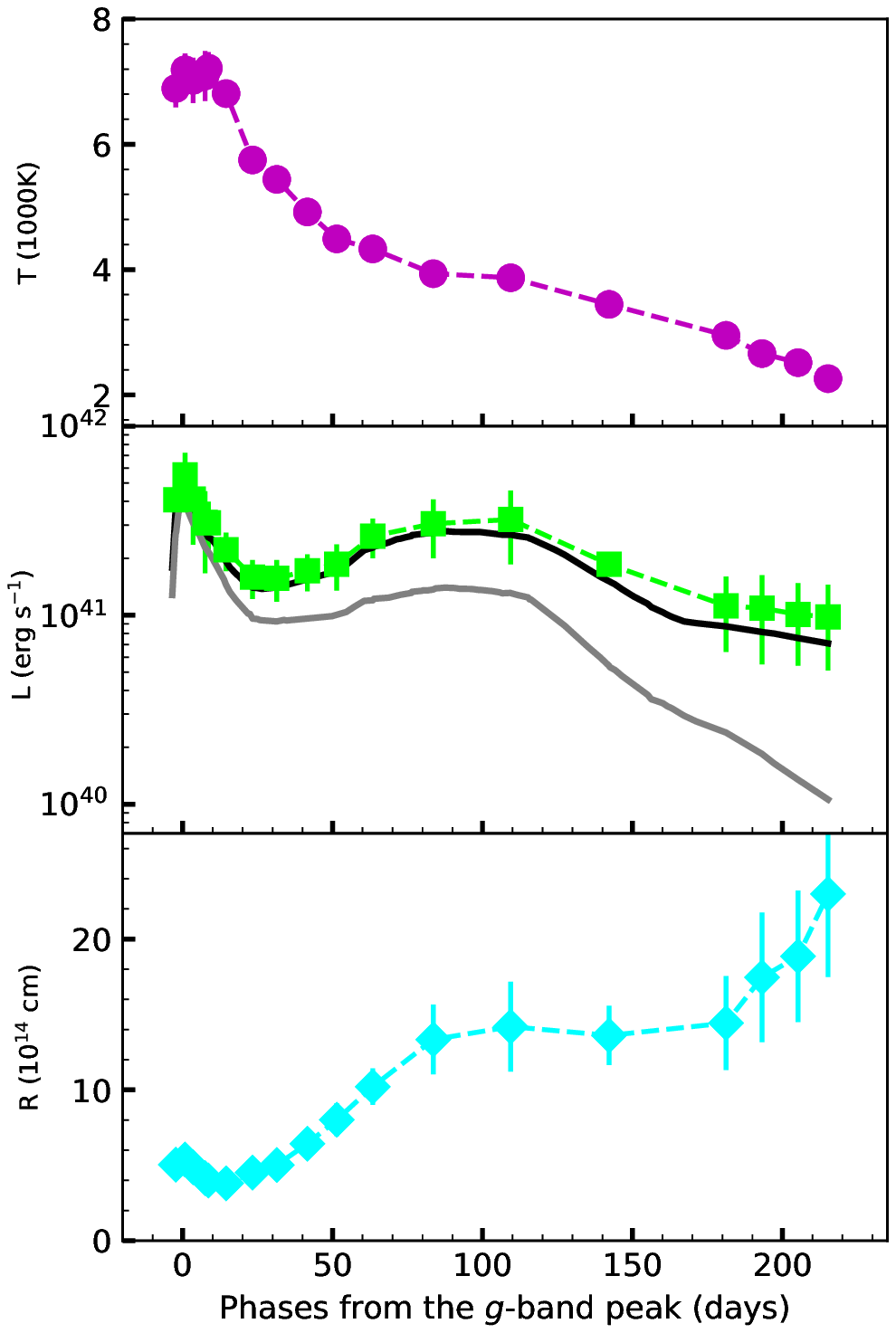}}
   \caption{Left: Evolution of the SED at some selected epochs spanning the entire evolution of \object{AT~2017jfs}.
 Top-right: Evolution of the effective temperature. Middle-right: Bolometric light curve of \object{AT~2017jfs} (green squares and dashed line), compared with
the {\sl uvoir} (black solid line) and  {\sl opt} (gray solid line) pseudo-bolometric curves (see text).
Bottom-right: Evolution of the photospheric radius  of \object{AT~2017jfs}.}
              \label{fig4}%
    \end{figure}

To this aim, the SED is computed for a few representative epochs, and the observed data are fitted by a single black-body function.
The early-time (near the blue maximum) SEDs do not contain $u$, $i$, $z$-band and NIR observations, while  $u$-band data are not available from
about two months after peak. Finally, the fluxes in the blue optical bands are not available at very late phases, because the object
was below the detection thresholds in those filters. The resulting black-body fits are shown in Fig. \ref{fig4} (left panel). 
Until about three months after maximum, observations are well modeled by black-body fits,
although from $\sim$3-4 weeks the line-blanketed $u$-band sits below the adopted models.
After the red peak, a single black body is not sufficient to accurately represent the observed SEDs in the blue region (see the inset in the left 
panel of Fig. \ref{fig4}). This happens when the NIR light curves of \object{AT~2017jfs} start a new rise before the object is in heliacal conjunction. 
This is possibly due to the contribution of a second black-body component peaking at longer wavelengths that cannot be properly fitted 
because of the inadequate wavelength coverage of our observed SED, in particular towards the mid- and far-infrared domains. The nature of 
this putative cold component is unclear. It is possibly due to an IR echo from distant pre-existing dust or, more likely,
to the condensation of newly formed dust, as suggested by the early appearance of molecular bands in the spectra and 
the strong blue-shift of the late H$\alpha$ emission (Section \ref{spectroscopy}).

The evolution of the effective temperature  is shown in the 
top-right panel of Fig. \ref{fig4}. The temperature remains roughly
constant at about 7000 K during the blue peak. Soon after maximum, the temperature declines very rapidly, reaching $\sim$4500 K 
at 50 d. Later on, the temperature fades more slowly, down to about 2300 K at 215 d, although this value is uncertain, as 
it was inferred from a poor, single black-body fit (see inset in Fig. \ref{fig4}, left panel).

The temporal evolution of the bolometric luminosity of \object{AT~2017jfs}, inferred by integrating the black-body fluxes over the entire wavelength range, 
 is shown in Fig. \ref{fig4} (mid-right panel), and is compared with the pseudo-bolometric curves obtained by accounting for the contribution
of the optical plus NIR bands ({\sl uvoir}), and the optical bands ({\sl opt}) only. For the first peak we obtain a bolometric luminosity $L_{bol} \sim 
5.5 \times 10^{41}$ erg s$^{-1}$, which is  comparable to those of other intermediate-luminosity optical transients 
\citep{ber09,sok12} or faint core-collapse SNe \citep{pasto04,spiro14}. After the post-peak decline (with a minimum of 
$L_{bol} \sim 1.5 \times 10^{41}$ erg s$^{-1}$), the bolometric light curve rises to the second peak with $L_{bol} \sim 3.2 \times 10^{41}$ erg s$^{-1}$, 
and then declines again. After $\sim$150 d, in coincidence with the late NIR brightening, the bolometric light curve flattens to $L_{bol} \sim10^{41}$ 
erg s$^{-1}$. We note that the bolometric luminosity of the blue peak in \object{AT~2017jfs} is twice that of the red peak.
This is a major difference with \object{NGC4490-2011OT1}, where the red peak was twice as luminous as the early blue peak \citep[see][their Fig. 11]{pasto19}. 
This discrepancy is likely a consequence of the large UV contribution during the blue peak that was not accounted for in the pseudo-bolometric light
curve of \object{NGC4490-2011OT1}.

Using the Stefan-Boltzmann law, with the luminosities and temperatures estimated above, we infer the evolution of the radius at the photosphere
for \object{AT~2017jfs} (Fig. \ref{fig4}, bottom-right panel). The radius $R$ at blue peak slightly exceeds $ 5 \times 10^{14}$ cm ($R \approx 7700~R_\sun$). 
After a modest decline, the radius rapidly increases reaching $\sim19000~R_\sun$ at about 80 d, and then remains roughly constant until $\sim6$ months.
During the last month of the monitoring campaign of \object{AT~2017jfs}, we observe a further fast increase in the photospheric radius, which exceeds
$33000~R_\sun$ at 215 d. This rise in the photospheric radius and the dramatic decline of the effective temperature at very late 
phases favor the formation of new dust, like in RN \object{V838~Mon} \citep{bond03}. This is also consistent with the blueshift
of the H$\alpha$ emission shown in Fig. \ref{fig2} (left).
 
\section{Discussion and conclusions}

\citet{pasto19} presented optical data for a wide sample of extra-galactic LRNe, all of them showing double-peaked light curves with maximum absolute magnitudes 
$M_V$ in the range $-12.5$ to $-15$ mag. They discuss the observational similarity of LRNe with fainter ($M_V>-10$ mag) RNe 
discovered in the Milky Way, and agree with \citet{koc14} and \citet{smi16} that all these transients are explained in a similar binary-interaction
framework. Most likely, they result from stellar merging events that occurred after the ejection of the common envelope. \citet{lip17} and \citet{mac17}
discuss the structured light curves of LRNe. A plausible scenario for the double-peak light curve of \object{AT~2017jfs} invokes an initial mass outflow as 
a consequence of the merging event, followed by a later interaction with the common envelope. This would produce the first luminosity peak and the
spectra resembling those of type-IIn SNe. During the second peak, the photospheric radius ($\sim2 \times 10^4~R_\sun$) is likely coincident with
that of the ejected common envelope. With the temperature decline, the H recombines, and the released radiation determines the broad red maximum.
According to \citet{met17}, the double-peak light curve of LRNe is explained with a modest mass ejection following the coalescence, with
the early peak being due to the release of thermal energy from the  fast ejecta in free expansion along the polar axes. The late red peak 
would result from shock-powered emission in the collision between the fast shell and pre-existing material in the equatorial plane. This  would also generate
a cool dense shell, which is an ideal site for late dust formation, as likely observed in \object{AT~2017jfs}.
\citet{bar14} provided a somewhat different interpretation. The rapid coalescence generates a violent forward shock which leads the photospheric temperature to largely increase,  
producing the blue light curve peak. This phase is followed by the fast adiabatic expansion of the envelope with thermal energy carried out with some delay to the outer layers
producing the broad red maximum.

\citet{koc14} proposed that the wide range of peak luminosities observed in RN/LRN events (over 4 orders of magnitudes in luminosity) is tightly connected 
with the total mass of the binary system, with faint RNe having progenitor systems of the order of $1~M_\sun$ and intermediate-luminosity events like 
\object{V838~Mon} of $\lesssim10~M_\odot$.
Luminous transients such as \object{AT~2017jfs}, \object{SNhunt248} \citep{mau15,kan15}, and \object{NGC~4490-2011OT1} \citep{smi16,pasto19} 
likely arise from more massive binaries \citep[up to $50-60~M_\sun$,][]{mau18}. While for \object{AT~2017jfs} we do not have any direct 
information on the progenitor system and the pre-outburst light-curve evolution, its luminous light curve would favor a massive binary
as precursor of \object{AT~2017jfs}.

A possible correlation between outflow velocities and light-curve peak luminosities for merger candidates is presented in \citet{mau18}, but
includes intermediate-luminosity red transients similar to \object{SN~2008S} and \object{M85-OT} whose nature is  debated \citep[e.g.,][]{bot09,kas11,kul07,pasto07}.
Since for \object{AT~2017jfs} we measure an expansion velocity
$v_{FWHM} \approx 700$ km s$^{-1}$ and log~$(L_{peak}/L_\sun) \approx 7.2$, the object is positioned very close to \object{NGC4490-2011OT1} in their Fig. 12, hence supporting
the parameters trend discussed in  \citet{mau18}.

While RNe from relatively low-mass stars are expected to be quite common, luminous events are 
extremely rare \citep{koc14}. In particular, following \citet[][their Fig. 3]{koc14}, AT~2017jfs-like events would occur at a rate of 
$<10^{-4}$ yr$^{-1}$ within 1 Mpc. Therefore, within a volume of 40 Mpc in radius, we should find about three events per year, 
which is roughly consistent with observations.
In fact, while we observed at least four RNe with $M_V\gtrsim-10$ mag in the Milky Way in the past two decades (\object{V4332~Sgr}, \object{V838~Mon}, \object{V1309~Sco} and \object{OGLE-2002-BLG-360}), 
LRNe brighter than $M_V < -10$ mag were never discovered in our Galaxy, with only less that ten objects observed within
40 Mpc in the past few years. 

Due to the limited number of objects discovered so far and incomplete data sets, RNe/LRNe are still not fully
understood. Well-sampled, multi-band light curves extending to longer wavelengths and high-S/N spectra with good resolution 
are essential tools for improving their characterization. Discovering new LRNe at larger distances and RNe outside the Local Group 
 is crucial for understanding the physics of these objects, and for providing reliable intrinsic rates. These are key objectives 
of the Large Synoptic Survey Telescope \citep{LSST09} and other future-generation surveys.

\begin{acknowledgements}
We thank Rubina Kotak for useful suggestions.
YZC is supported by the China Scholarship Council (No. 201606040170).
MF is supported by a Royal Society - Science Foundation Ireland University Research Fellowship. 
NER acknowledges support from the Spanish MICINN grant ESP2017-82674-R and FEDER funds.
S.Bose, PC and SD acknowledge Project 11573003 supported by NSFC. This research uses data obtained through the Telescope Access Program (TAP), which has been funded by the National Astronomical Observatories of China, the Chinese Academy of Sciences, and the Special Fund for Astronomy from the Ministry of Finance. 
S.Benetti is partially supported by PRIN-INAF 2017 ''Toward the SKA and CTA era: discovery, localization, and physics of transient sources.'' (PI: M. Giroletti). 
KM acknowledges support from STFC (ST/M005348/1) and from H2020 through an ERC Starting Grant (758638).
AF acknowledges the support of an ESO Studentship. 
AMT acknowledges the support from the Program of development of M.V. Lomonosov Moscow State University (Leading Scientific School ``Physics of stars, relativistic objects and galaxies''.
CT, AdUP, DAK and LI acknowledge support from the Spanish research project AYA2017-89384-P, and from the “Center of Excellence Severo Ochoa” award for the IAA (SEV-2017-0709). CT and AdUP acknowledge support from funding associated to Ram\'on y Cajal fellowships (RyC-2012-09984 and RyC-2012-09975). DAK and LI acknowledge support from funding associated to Juan de la Cierva Incorporaci\'on fellowships (IJCI-2015-26153 and IJCI-2016-30940).

The Pan-STARRS1 Surveys (PS1) have been made possible through contributions of the Institute for Astronomy, the University of Hawaii, the Pan-STARRS Project Office, the Max-Planck 
Society and its participating institutes, the Max Planck Institute for Astronomy, Heidelberg and the Max Planck Institute for Extraterrestrial Physics, Garching, The Johns Hopkins 
University, Durham University, the University of Edinburgh, Queen's University Belfast, the Harvard-Smithsonian Center for Astrophysics, the Las Cumbres Observatory Global Telescope 
Network Incorporated, the National Central University of Taiwan, STScI, NASA under Grant No. NNX08AR22G issued through the Planetary Science Division of the NASA Science 
Mission Directorate, the US NSF under Grant No. AST-1238877, the University of Maryland, and Eotvos Lorand University (ELTE). Operation of the Pan-STARRS1 telescope is supported 
by NASA under Grant No. NNX12AR65G and Grant No. NNX14AM74G issued through the NEO Observation Program.  
 This paper is also based upon work supported by AURA through the National Science Foundation under AURA Cooperative Agreement AST 0132798 as amended.
ATLAS observations were supported by NASA grant NN12AR55G. 
NUTS is supported in part by the Instrument Center for Danish Astrophysics (IDA). 

This work is based on observations collected at the European Organisation for Astronomical Research in the Southern Hemisphere under ESO programmes 199.D-0143(G,I,K,L). 
This work makes use of observations from the LCOGT network. 
It as also based on observations made with the 2.2m MPG telescope at the La Silla Observatory, the Nordic Optical Telescope (NOT), operated on the island of La Palma jointly by Denmark, Finland, Iceland, 
Norway, and Sweden, in the Spanish Observatorio del Roque de los Muchachos of the Instituto de Astrof\'isica de Canarias; 
the 1.82~m Copernico Telescope of INAF-Asiago Observatory; the Gran Telescopio Canarias (GTC), installed in the Spanish Observatorio del Roque  
de los Muchachos of the Instituto de Astrof\'isica de Canarias, in the Island of La Palma;  
the Liverpool Telescope operated on the island of La Palma by Liverpool John Moores University at the Spanish Observatorio del Roque de los Muchachos of the Instituto de Astrof\'isica 
de Canarias with financial support from the UK Science and Technology Facilities Council; the 6m Big Telescope Alt-azimuth and the Zeiss-1000 Telescope of the Special Astrophysical Observatory, 
Russian Academy of Sciences. 

We thank Las Cumbres Observatory and its staff for their continued support 
of ASAS-SN. ASAS-SN is supported by the Gordon and Betty Moore Foundation through grant GBMF5490 to the Ohio State University and NSF grant AST-1515927. Development of ASAS-SN has  been supported by NSF grant AST-0908816, the Mt. Cuba Astronomical Foundation, the Center for Cosmology and AstroParticle Physics at the Ohio State University, the Chinese Academy of Sciences South America Center for Astronomy (CAS-SACA), the Villum Foundation, and George Skestos.

\end{acknowledgements}

\begin{appendix}
\section{Additional data}

   \begin{table*}[!t]
      \caption[]{General information on the spectra of \object{AT~2017jfs}. The phases are from the $g$-band maximum.}
         \label{tab}
     $$         \begin{array}{lllllll}
            \hline  \hline
            \noalign{\smallskip}
    $Date$ &  $MJD$ & $Phase~(days$) & $Instrumental~configuration$ & $Exptime~(s)$ & $Res~(\AA)$ & $Range~(\AA)$ \\ \hline
$2018~Jan~6 $ & 58124.31 & +9.5   & $NTT+EFOSC2+gm13         $ & 600                   & 18      &$ 3650-9200 $\\
$2018~Jan~10$ & 58128.26 & +13.5  & $NOT+ALFOSC+gm4          $ & 3600                  & 18      &$ 3600-9650 $\\ 
$2018~Jan~17$ & 58135.33 & +20.5  & $NTT+EFOSC2+gm11         $ & 2\times3600           & 14      &$ 3350-7450 $\\
$2018~Jan~20$ & 58138.16 & +23.4  & $NOT+ALFOSC+gm4          $ & 3600                  & 14      &$ 3500-9700 $\\
$2018~Jan~24$ & 58142.41 & +27.6  & $P200+DBSP+gt316/7500    $ & 1200                  & 5.5     &$ 5400-9500 $\\
$2018~Feb~13$ & 58162.22 & +47.4  & $NTT+EFOSC2+gm11         $ & 2\times3600           & 14      &$ 3350-7450 $\\
$2018~Feb~23$ & 58172.10 & +57.3  & $GTC+OSIRIS+R1000B+R1000R$ & 1800+1800             & 7.0+7.8 &$ 3850-9250 $\\
$2018~Mar~20$ & 58197.10 & +82.3  & $GTC+OSIRIS+R1000B+R1000R$ & 1800+1800             & 7.0+7.8 &$ 3650-9350$\\
$2018~Apr~22$ & 58230.94 & +116.1 & $BTA+SCORPIO+VPHG550G    $ & 2\times1200           & 13      &$ 3650-7900 $\\
$2018~Apr~29$ & 58237.05 & +122.3 & $GTC+OSIRIS+R1000B+R1000R$ & 2\times900+2\times900 & 7.0+7.8 &$ 3650-9200 $\\
$2018~May~23$ & 58261.06 & +146.3 & $GTC+OSIRIS+R1000B+R1000R$ & 2\times1600+1600      & 7.0+7.8 &$ 3650-9250 $\\
$2018~Jun~1 $ & 58270.98 & +156.2 & $GTC+OSIRIS+R1000R       $ & 2\times1800           & 7.8     &$ 5100-9350 $\\
$2018~Jun~2 $ & 58271.97 & +157.2 & $GTC+OSIRIS+R1000R       $ & 2\times1800           & 7.8     &$ 5100-9350 $\\
$2018~Jun~16$ & 58285.93 & +171.1 & $GTC+OSIRIS+R1000R       $ & 4\times1500           & 7.8     &$ 5100-9250 $\\
\hline                    
        \noalign{\smallskip}
         \end{array}
     $$ 
\tablefoot{
    NTT = 3.58 m New Technology Telescope (ESO-La Silla, Chile);\\
    NOT = 2.56 m Nordic Optical Telescope (La Palma, Canary Islands, Spain);\\
    P200 = 5.1 m (200-inch) Hale Telescope (Mt. Palomar, California, USA);\\ 
    GTC = 10.4 m Gran Telescopio Canarias (La Palma, Canary Islands, Spain);\\
    BTA = 6.05m Bolshoi Teleskop Alt-azimutalnyi (Special Astrophysical Observatory, Karachay-Cherkessian Republic, Russia).}
   \end{table*}

\end{appendix}

\end{document}